# Diffusion of Mn in gallium nitride: Experiment and modelling

Rafal Jakiela[1], Katarzyna Gas[1,2], Maciej Sawicki[1], Adam Barcz[1,3]

[1]Institute of Physics, Polish Academy of Sciences, Aleja Lotnikow 32/46, PL-02668 Warsaw, Poland
[2]Institute of Experimental Physics, University of Wrocław, Pl. Maxa Borna 9, 50-204 Wrocław, Poland
[3]Institute of Electron Technology, Lotników 32/46, Warsaw, Poland

**Abstract**

The control over the structural homogeneity is of paramount importance for ternary nitride compounds – the second most important semiconducting material-class after Si, due to its unrivalled applicability in optoelectronics, and high power/high frequency electronics. Therefore it is timely to investigate possible mechanisms influencing the crystallographic constitution of the material. In this work the diffusion mechanism of manganese in gallium nitride is investigated in two types of epilayers: Mn-implanted metalorganic vapour phase epitaxy grown GaN and (Ga,Mn)N solid solution grown by molecular beam epitaxy. The extent of the Mn diffusion is established by secondary ion mass spectrometry. Analysis of the Mn profiles in the implanted samples in the frame of the infinite source diffusion led to the establishment for the first time of the pre-exponential factor $D_O = 2\times10^{-4}$ cm$^2$/s and diffusion activation energy $E_A = 1.8$ eV describing the diffusion coefficient of Mn in GaN. Modelling of the out-diffusion of Mn from (Ga,Mn)N layers based on these values in turn allows to provide an explanation of the origin of the ubiquitously observed near-the-surface sizable depletion of Mn in (Ga,Mn)N, resulting from the blocking of the Mn out-diffusion by Mn-oxide and the correspondingly formed space-charge layer on the surface.

## 1. Introduction

Nitride family of wide gap semiconducting compounds, of which GaN is the most representative member, has shown a tremendous potential for high power, high frequency communications and photonic applications, reaching now the status of the second technologically most important semiconductor after Si [1], so the prospect of spintronics application of GaN have gained in significance [2]. However, the most obvious way of magnetic functionalization of GaN by alloying it with transition metals (TM), mainly Mn, has not led to the expected technological breakthrough [3]. As it was pointed out Mn (and other transition metals) *d*-levels provide not only localized spins but also they contribute to the cohesive energy of the crystal what frequently leads to aggregation [4,5]. In this case, driven either by a spinodal decomposition (mostly) or by nucleation mechanisms (binodal decomposition, less frequently) regions with high concentrations of magnetic constituents, either commensurate with the host lattice (chemical phase separation) or in the form of precipitates (crystallographic phase separation) are formed in the otherwise nonmagnetic matrix. An important progress in the physical insight into the issue has been made after realization that the incorporation and distribution of magnetic ions depend sizably on codoping by shallow impurities or electrically active defects [6], what provides an efficient control over the aggregation of magnetic cations either at the growth surface or in the film volume [7-9]. On the other hand, extremely high levels of compensating donors in homogenously doped samples change the oxidation state of the magnetic ions what affects the type of their magnetic coupling. In the case of GaN with Mn, the *n*-type doping leads to $Mn^{2+}$ configuration for which spin-spin coupling is antiferromagnetic [10,11]. In the compensation–free films containing dilute $Mn^{3+}$ [12-14], the mid band gap located $Mn^{2+/3+}$ acceptor level [15,16] precludes the existence of carrier-mediated spin-spin coupling. However, in this case the short-range superexchange Mn-Mn interaction results in homogenous ferromagnetic (FM) coupling [14, 17-20] characterized by a mediocre magnitude of Curie temperature, not exceeding 13 K even for Mn content *x* as high as 10%, despite exhibiting the highest saturation magnetization in the whole dilute ferromagnetic semiconductors family [19]. Nevertheless, this relative unique realization of ferromagnetism in the insulating semiconductor [13, 21, 22] has opened prospects for studies of a dissipation-less information transfer [23], or in a combination with piezoelectricity of this inversion-symmetry-lacking wurtzite structure material, a direct electrical control of $Mn^{3+}$ ions magnetization has been shown possible [24]. Yet, a paramount amount of growth-related issues remain to be addressed in this material system. Mainly, they are related to a low Mn incorporation rate, which sizable improvement is essential to expand Curie temperature range towards the commercially important ranges. Other important topics aim at identifications of the sources of an uneven Mn distribution within the material, which leaves a great impact on the magnetic uniformity [14]. Our studies let us to point out one more, previously not considered in the context of material structural constitution, mechanism, namely, the high diffusivity of transition metal species during the growth and cool down process which takes place at temperatures at which forced by the nonequilibrium growth condition concentration of TM species exceeds by far the corresponding solubility limit. An issue particularly relevant for the two main non-equilibrium growth methods: metalorganic vapour phase epitaxy (MOVPE) and molecular beam epitaxy (MBE) which require very high growth temperatures $T_g$: 800 – 900°C [12] and 600 – 760°C [14,17,19], respectively.

Despite its importance, surprisingly, diffusion of Mn in GaN has not received much attention in the literature so far. Implantation and subsequent annealing at 1050°C for 12 min. were performed in Ref. (25) and subsequent Rutherford backscattering spectrometry (RBS) investigations showed no significant redistribution of implanted atoms. Authors of Ref. (26) investigated Mn-doped GaN obtained by diffusion from an Mn layer deposited on the sample's surface and annealed at 500 or 800°C for 6 h. They showed that the Mn diffusion range was either 17 or 25 nm, respectively. The diffusion method was also used in Ref. (27). In this case the secondary ion mass spectrometry (SIMS) profiling showed clearly Mn diffusion into GaN but the corresponding diffusion coefficient was not extracted from the experiment.

In this study we experimentally investigate and numerically quantify Mn atoms diffusion-related processes in: (i) Mn implanted and subsequently annealed GaN, and (ii) MBE grown (Ga,Mn)N layers cooled down from $T_g$ to the room temperature in a controlled manner. The paper is organized as follows. Firstly, after establishing the Mn solubility limit in GaN, the pre-exponential factor and diffusion activation energy describing the diffusion coefficient of Mn in GaN are established on the account of the diffusion from the infinite source model. These values are used further in a quantitative account of the out-diffusion of Mn from epitaxially grown (Ga,Mn)N layers allowing to put forward a plausible explanation of the origin of the ubiquitously observed near-the-surface sizable depletion of Mn in (Ga,Mn)N and other structural and compositional nonuniformities observed in epitaxially grown dilute magnetic semiconductors.

## 2. Experimental details

There are two methods of sample preparation for this study. In the first one, an MOVPE grown GaN layer on the *c*-plane sapphire substrate has been Mn implanted at room temperature with a dose of $10^{16}$ cm$^{-2}$ at the energy of 150 keV. Stopping and range of ions in matter (SRIM) simulation of implantation situates the maximum of Mn concentration at a depth of 95 nm, which agrees with the SIMS depth profile of the as-implanted sample presented in Figure 1. The post-implantation annealing has been performed in the same MOVPE reactor at 900, 1000, and 1100°C for 20 or 40 minutes at each temperature under $N_2$ atmospheric pressure.
The out-diffusion of Mn from (Ga,Mn)N layers is studied in samples grown by MBE method in a Scienta-Omicron Pro-100 MBE chamber equipped with a radio-frequency nitrogen plasma source. Commercially available 3 µm thick GaN(0001) template layers deposited on *c*-oriented 2" sapphire have served as substrates. The beam equivalent pressures of Mn and Ga were set to $0.5 \times 10^{-7}$ and $2 \times 10^{-7}$ mbar, respectively. A nitrogen flux of 1.3 standard cubic centimetres per minute at a plasma power of 400 W is used to provide reactive nitrogen. The growth temperature is set to 605, 660 and 700°C, what results in different Mn incorporation of about $2.6\times10^{20}$, $1\times10^{21}$ and $3.7\times10^{21}$ atoms/cm$^3$ ($x$ = 0.6, 2.2 and 8.4%) as determined by SIMS and confirmed by high resolution x-ray diffraction, and magnetic saturation at low temperatures [14]. Also the actual growth temperature affects the thickness of the obtained layers, so it amounts to 40, 70, and 95 nm despite the same growth time of 2 h. Additional detailed structural and magnetic characterization of this set of samples is given elsewhere [14]. After completing the growth the samples are cooled down to about 100°C with a cooling rate of 50°C/min, that is the cooling time amounts to about 600, 670, and 720 s, respectively. One more sample has been grown at 620°C at the same conditions as described above and additionally post-growth annealed by keeping it for 30 min at $T_g$ and subsequently cooled down with the same cooling rate. The resulting Mn profiles in the samples are established with SIMS technique using a CAMECA IMS6F microanalyzer. For implanted samples, the SIMS measurement is performed with oxygen (O$_2$+) primary beam at the energy of 8 keV with the current kept at 70 nA. For MBE samples, oxygen (O$_4$+) primary beam at the 2 nA current is used to increase the depth resolution in the subsurface region of the (Ga,Mn)N films, what resulted in a somewhat reduced sputter rate. The size of the raster is about 150 × 150 microns and the secondary ions are collected from a central region of approximately 60 microns in diameter. Magnitudes of Mn concentrations are derived from the intensity of Mn+ species with the matrix signal N+ taken as a reference. A non-annealed fragment of Mn-implanted GaN sample serves as a calibration standard.

## 3. Results and discussion

We start the data analysis from a comparison of Mn depth profiles of the as-implanted GaN layer and its maximally annealed counterpart that is annealed at 1100°C for 40 min. As shown later these condition results in the deepest diffusion, actually reaching the end of the 3 µm thick layer. The data presented in Figure 1 indicate that: (i) the solid solubility limit of Mn in GaN at 1100°C is as low as approximately $2\times10^{17}$ at/cm$^3$, and (ii) the implanted near-the-surface part of the layer has hardly changed during annealing what indicates that the top ~0.2 µm of the layer can be treated as an infinite source, and (iii) the detection limit in this experiment is as low as $5\times10^{15}$ at/cm$^3$.

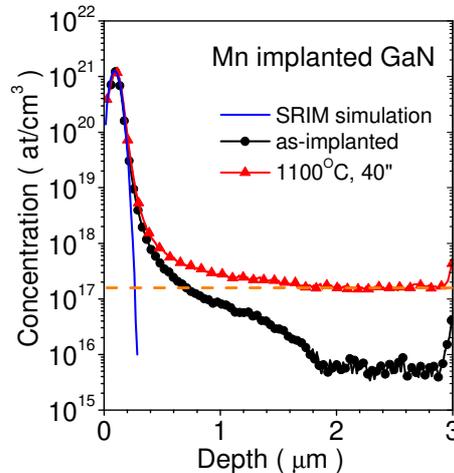

**Figure 1.** (points) Mn depth profiles in as-implanted GaN samples established by secondary ion mass spectrometry. Solid line represents results of stopping and range of ions in matter (SRIM) simulation relevant to the current implantation conditions.

Since we can treat this experiment as a diffusion from an infinite source the diffusion profiles should be described by complementary error function – *erfc*:

$$C(z,t) = C_S \, erfc\left(\frac{z}{\sqrt{4Dt}}\right) \quad (1)$$

Where: $C_S$ is the maximum concentration of diffused species, which corresponds to the surface concentration in the case of the infinite source experiment; $D$ is the diffusion coefficient; $t$ is the time of annealing; and $z$ measures the depth from the surface.

In the standard case of the diffusion from the infinite source, the diffused element is deposited on the sample's surface and the studied sample initially contains no diffused species. In the case of implanted samples we start with an already existing initial $z$ profile in the sample. As indicated in Figure 1 it takes nearly 2 µm for, $x$ to decrease smoothly from about $10^{21}$ to the detection limit of $5 \times 10^{15}$ at/cm$^3$. Therefore to extract the real profile related to the annealing-driven diffusion we have to subtract the profile of the as–implanted layer, serving now as the reference profile, from the profiles obtained after annealing. The Mn profiles obtained this way, as well as the fitted *erfc* functions are shown in Figure 2. The magnitudes of the diffusion coefficients obtained upon fitting of Eq. (1) change substantially for these temperatures and range from $4 \times 10^{-12}$ to $1 \times 10^{-10}$ cm$^2$/s as the annealing condition change from 900°C and 40 min to 1100°C and 20 min, respectively. All the established values of $D$ are shown on the Arrhenius plot in panel c) of Figure 2. We add that practically no diffusion takes place for annealing lasting only 20 min at 900°C and the 40 min annealing process at 1100°C gives practically a flat $z$-profile, shown already in Figure 1.

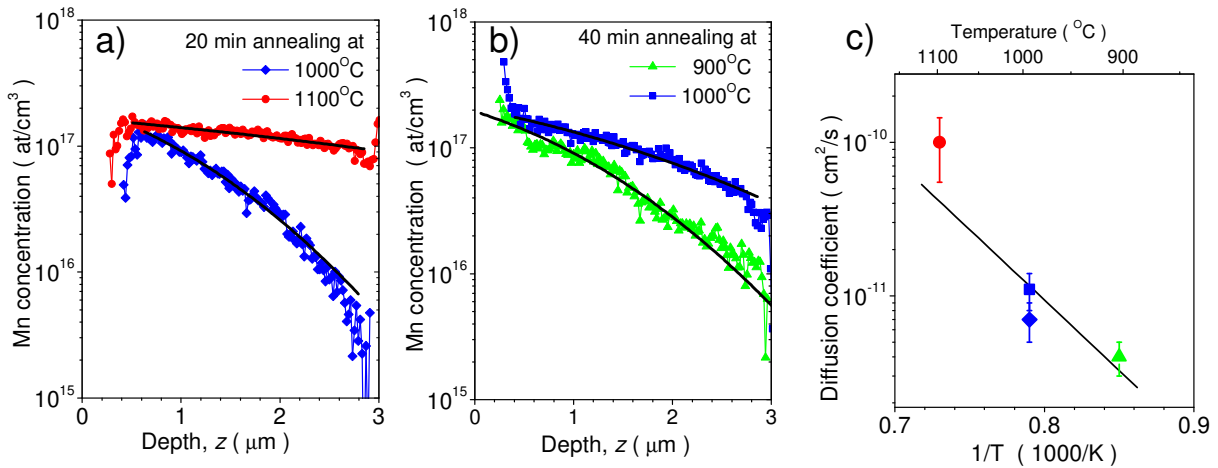

**Figure 2. (a-b)** Mn concentration profiles (symbols) in implanted GaN after a) 20 min., b) 40 min. annealing at the respective temperatures. The solid black lines represent fitted Eq. 1. **c)** Arrhenius plot for Mn bulk diffusion in GaN (points established upon the fitting of the data in the previous panels). The black line represents the fit of the classical Arrhenius equation (Eq. 2).

The data collected in Figure 2c permit us to establish both the temperature independent pre-exponent factor $D_O = 2 \pm 0.2 \times 10^{-4}$ cm$^2$/s and the activation energy for the diffusion process $E_A = 1.8 \pm 0.1$ eV using the classical Arrhenius equation:

$$D = D_O \exp\left(\frac{-E_A}{kT}\right) \quad (2)$$

Such a magnitude of $E_A$ rules out the kick-out mechanism in which Mn atoms move Ga into an interstitial positions because of a high, over 10 eV, formation energy of interstitial Ga [28]. It is also too high for a pure interstitial mechanism. Energy barrier for this diffusion process is over 1 eV lower than in the case of Mn interstitial diffusion in gallium arsenide [29].

However, our findings are in favor of an interstitial-substitutional diffusion (ISD) mechanism, so a sufficient concentration of Ga vacancies $V_{Ga}$ is needed. For a system in thermodynamic equilibrium the distribution and concentration of impurity species are determined by their formation energy. In the case of epitaxial GaN a typically substantial amount of residual O and Si donors guaranties ample concentration of $V_{Ga}$, because of their very low formation energy of ~0.2 eV in *n*-type GaN [28]. Importantly, the established here $E_A$ for Mn compares very favourably to that obtained for Si diffusion, exhibiting also ISD character [30].

The second analyzed here phenomenon is Mn out-diffusion from MBE grown samples. Interestingly, this previously disregarded mechanism is ubiquitously observed in (Ga,Mn)N [14, 19, 24, 31-33] as a characteristic a few tens of nm wide drop of Mn concentration observed exclusively near the surface, as exemplified in Figure 3, for the studied here MBE grown layers.

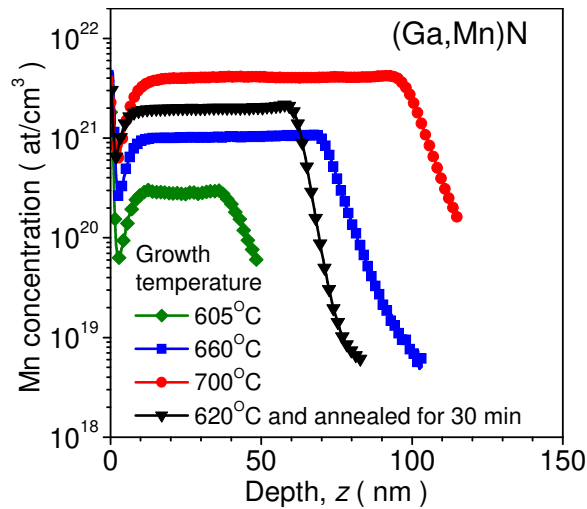

**Figure 3.** Manganese depth profiles in molecular beam epitaxy grown (Ga,Mn)N layers as measured by secondary ions mass spectrometry.

The behaviour of a dopant in a foreign material is determined primarily by two quantities: diffusivity and solid solubility. Both generally increase with temperature. Because of this the host is capable to accept more doping atoms at an elevated temperature than at the base one, e.g. room temperature. Therefore during the high temperature epitaxy, typically 600 - 700°C for (Ga,Mn)N growth by in MBE or 900 - 1000°C for MOVPE process, a greater number of Mn atoms are incorporated into the GaN lattice than this lattice can contain at room temperature. This excess of Mn species has to be expelled by the host during the post-growth cool-down. Experimental evidence, mainly magnetic studies on a wide range of samples for which no high temperature ferromagnetic-like signal was observed [12-14, 17, 18] indicate that Mn diffusion in GaN is a more probable channel than agglomeration in some kind of nanocrystals. However, the existence of the free surface, where the arriving Mn species can meet oxygen atoms and passivate, provides the necessary driving force directed towards the surface.

In the classical time-honored case of $SiO_2$/Si [34] where the direction of diffusion (in- or out-) depends on the magnitudes of solubility limit of oxygen at a given temperature $C_S$ and on the actual oxygen concentration in silicon $C_B$ the resulting profile is described by:

$$C(z,t) = C_S + (C_B - C_S) erf\left(\frac{z}{\sqrt{4Dt}}\right) \quad (3)$$

It follows that at $T$ where $C_S > C_B$ the given element diffuses from the surface into the bulk, and when $T$ is such that $C_S < C_B$ the foreign atoms outdiffuse. And this is the case that we are dealing here with. The before established $C_S$ of Mn in GaN = $2 \times 10^{17}$ cm$^{-3}$ is orders of magnitudes less than the value of $C_B$ ($C_B = 0.5xN_0$, where $N_0 = 4.4 \times 10^{22}$ cm$^{-3}$ is the cation concentration in GaN) reached upon the nonequilibrium growth process. Experimentally, we do not observe the required by Eq. 3 drop of Mn concentration down to $C_B$ on the surface due to the depth resolution of our SIMS measurements which we evaluate to be at the best 3 nm.

However, it has to be added that chemical reaction of Mn with oxygen at the surface can but does not have to inhibit the Mn outdiffusion for GaN. In the classical case of $SiO_2$/Si it is the $T$-driven direction of diffusion which leads either to rebuilding (outdiffusion) or to dissociation (in-diffusion) of the $SiO_2$ layer on Si. In the case of (Ga,Mn)N, since Mn is the diffusing species, it needs an external source of oxygen to be passivated after reaching the surface. Our conjecture is that a thick enough oxide layer may so efficiently hinder oxygen supply form the ambient that the arriving Mn species will form a space-charge layer at the GaN surface which will counteract the diffusion towards the surface. Since the diffusion is temperature- and time-dependent there is only a narrow part of the layer from which the Mn atoms can reach the surface before the space-charge layer is formed, what results in the Mn depletion at the near-the-surface part of the layer.

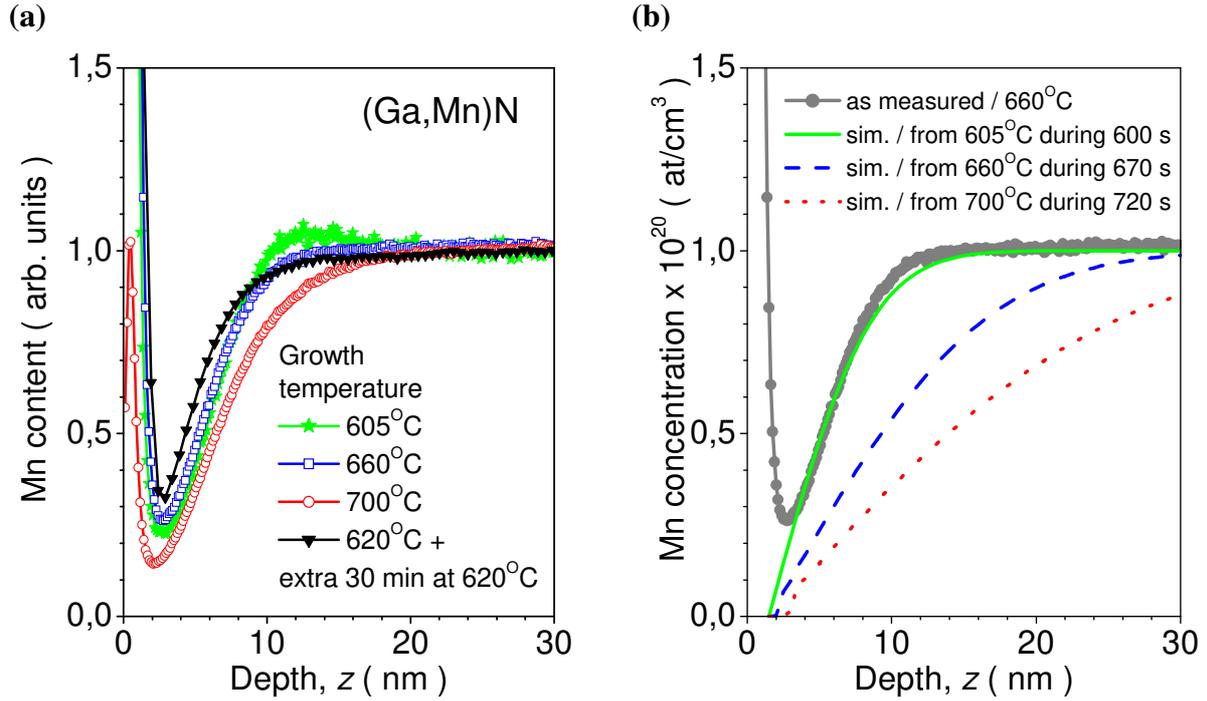

**Figure 4**. **(a)** Normalized Mn profiles of all the (Ga,Mn)N samples from this study. **(b)** (Grey bullets) Manganese depth profile for (Ga,Mn)N samples grown at the lowest growth temperature of 605°C. (Lines) Simulations of room temperature Mn profiles after cooling (Ga,Mn)N from 605, 660 and 700°C at 50 deg/min cooling rate taking the pre-exponential factor $D_O = 2\times 10^{-4}$ cm$^2$/s and diffusion activation energy $E_A = 1.8$ eV. The profiles are based on Eq. 4.

Now, having evaluated the Mn diffusion coefficient in GaN we attempt to give an account of the aforementioned "depletion effect". We start from noting, as experimentally established on the account of the Mn profiles presented in Figure 4(a), that the size of the depletion is concentration and growth temperature independent within our experimental accuracy, and that the whole process corresponds to the out-diffusion from an infinite source. Therefore, we employ the bare error function *erf*, which is fed with the established above values of $D_O$ and $E_A$ to describe $D(T)$. The whole function assumes the form given below:

$$C(z,t) = C_B \, erf\left(\frac{z}{\sqrt{4Dt}}\right) \qquad (4)$$

Since the Mn diffusion coefficient decreases respectively to the sample's temperature during the sample cooling, the diffusion coefficient is changed accordingly to the sample's cooling rate until the final temperature is reached. The relevant iteration is done using Forward Time Center Space (FTCS) method [35]. The calculated profiles, exemplified in Figure 4(b), indicate that as the result of the increased growth temperature, and so the associated with it thermal budget of the sample after its growth is terminated, the near-the-surface Mn depletion should deepen. But this is in a stark contrast with the experimental Mn profiles presented in Figure 4(a), particularly with that one of the additionally annealed sample at $T_g$, for which the sizably increased thermal budget should results in the strongest outdiffusion under the process modelled here. Such a discrepancy can be accounted for only when either the Mn diffusion process is characterized by a very low magnitude $E_A$ or the diffusion process is halted by another reason. Since the experimentally established value of $E_A$ is not sufficiently small, we opt here for the latter case in which the movement of Mn species towards the surface is impeded by the formation of the repelling space-charge layer related to the accumulated charged Mn species.

**Conclusions**

In this study the mechanisms of Mn diffusion in GaN have been experimentally investigated and theoretically modelled. In result for the first time the temperature dependence of the Mn diffusion has been accounted for and the relevant parameters of this dependency are established. In particular, the relatively low magnitude of the activation energy $E_A$ = 1.8 eV of the diffusion process points to the interstitial-substitutional diffusion mechanism through Ga vacancies. This in turn allows giving a quantitative account for the out-diffusion of Mn from epitaxially grown (Ga,Mn)N layers, providing a solid ground for a plausible explanation of the origin of the ubiquitously observed near-the-surface sizable depletion of Mn in (Ga,Mn)N. It has been argued that independently of the growth temperature and the Mn content its out-diffusion is blocked by the surface located Mn-oxide and the corresponding space-charge layer.

The experimental findings of this study, in particular, point out that surface-directed diffusion of the alien atoms (ions) in the host driven by their chemical activity at the surface, particularly effective at the growth temperature and during the cool down of the just-grown material, plays a decisive role in its final structural constitution. It constitutes therefore a new, and previously disregarded effect, which needs to be taken into account when details of the nanometer scale structures or devices are considered. Apart from the case detailed here of Mn in GaN, other examples in dilute magnetic semiconductor family include, for example, the formation of α-Fe nanocrystals in a proximity to the surface in un-capped GaN:Fe [36,37] and the presence of a sizable depth-dependence of Mn interstitial concentration in (Ga,Mn)As [38].

**Acknowledgements**

This study has been supported by the National Science Centre (Poland) through FUGA (DEC-2014/12/S/ST3/00549) and OPUS (DEC-2013/09/B/ST3/04175) and by the EC through the InTechFun (POIG.01.03.01-00-159/08) grants.